\documentclass[graybox]{svmult}

\usepackage{mathptmx}       
\usepackage{helvet}         
\usepackage{courier}        
\usepackage{type1cm}        
\usepackage{makeidx}         
\usepackage{graphicx}        
\usepackage{multicol}        
\usepackage[bottom]{footmisc}
\usepackage{epsfig}

\makeindex             
\usepackage{graphicx}
\usepackage{epsfig}
\usepackage{amssymb}

\usepackage{showlabels}

\begin{document}

\title*{The low  temperature corrections to the Casimir force between a sphere and a plane}
\author{Michael Bordag and Irina G. Pirozhenko}
\institute{Michael Bordag \at
Leipzig University,
Institute for Theoretical Physics,  04109 Leipzig, Germany,
\email{bordag@itp.uni-leipzig.de} \and Irina G. Pirozhenko \at
Bogoliubov Laboratory of Theoretical Physics, Joint Institute for
Nuclear Research and Dubna International University, Dubna 141 980,
Russia, \email{pirozhen@theor.jinr.ru}}
%
%
\maketitle

\abstract{We calculate the low temperature corrections to the
free energy for a sphere in front of a plane. First, the scalar field
obeying Dirichet or Neumann boundary conditions is considered.
Second, the electromagnetic field is studied, the sphere being
perfectly conducting and being a dielectric ball with both, constant permittivity and permittivity of the plasma model.}

\section{Introduction}
During the past decade significant attention was payed to the Casimir effect at finite temperature. The interest was triggered by the desire to measure the temperature dependent part of the force (see \cite{Klim09-81-1827}, section 4.D, for a review) and by the conceptual problems arising for $T\to0$ with some thermodynamic quantities (see \cite{Klim09-81-1827}, section 2.D).
Recent interest came also from the interplay between temperature and geometry investigated in \cite{gies10-25-2279,webe1003.3420} using world line methods. For a scalar field with Dirichlet boundary conditions on the interacting surfaces, a generic behavior $\sim T^4$ of the temperature dependent part of the force was found and attributed to open geometry.

In our previous paper \cite{bord10-81-085023} we investigated the interaction between a ball and a plane for both, the scalar and the electromagnetic fields. We used the exact functional determinant method (also called scattering approach or
'TGTG'-formula) and focused on the limit of small separation. We showed that the Proximity Force Approximation (PFA) is reproduced exactly for medium and high temperature. For low temperature, the leading order of the free energy is the vacuum energy (for which the PFA is reproduced, of course) and the temperature dependent part is a small addendum for which the PFA does not hold.

In the present paper we calculate the leading behavior of the temperature dependent part $\Delta_T\cal F$ of the free energy and  that of the force, $\Delta_T  f$,  for $T\to0$. Again, we consider the scalar and the electromagnetic fields. For the latter we allow for dispersion including fixed permittivity   and  permittivity following from the plasma model (see below, eq. (\ref{epplasma})). We use the functional determinant method and truncate the orbital momentum sum at some finite $l_m$. It turns out that the limit of $l_m\to\infty$ shows a sensitive dependence on the separation $a$ for $a\to0$.
Due to the space limitations we will be quite brief concerning technical details. These will be given in a separate paper.

We mention that the interplay of geometry and temperature in sphere-plane geometry was also studied in
\cite{Cana10-104-040403} and related papers using the scattering approach. In a numerical analysis the ambient temperature was fixed, while the radius of the sphere and plasma frequency varied.

Throughout the paper we use units with $\hbar=c=1$.

\section{The free energy at finite temperature}
In this section we collect the basic formulas for the free energy in functional determinant representation at finite temperature. We follow closely the notations used in \cite{bord10-81-085023}. At finite temperature, the free energy is given by
\begin{equation}\label{1.F1} {\cal F}=\frac{T}{2}\sum_{n=-\infty}^{\infty} {\rm Tr} \ln
\left(1-\mathbf{M}(\xi_n)\right),
\end{equation}
where $\xi_n=2\pi nT$ are the Matsubara frequencies. The matrix $\mathbf{M}$ results from the
scattering on the sphere and will be described below together with the meaning of the
trace. The sum over the Matsubara frequencies can be transformed into integrals using
the well known Abel-Plana formula. The free energy separates into two pieces,
\begin{equation}\label{1.FT} {\cal F}=E_0+\Delta_T {\cal F},
\end{equation}
where
\begin{equation}\label{1.E0} E_0=\frac12 \int_{-\infty}^{\infty}\frac{d\xi}{2\pi} \  {\rm Tr} \,\ln
\left(1-\mathbf{M}(\xi)\right)
\end{equation}
is the vacuum energy, i.e., the free energy at zero temperature, and
\begin{equation}\label{1.F2}  \Delta_T {\cal F}  =
\frac{1}{2}\int_{-\infty}^\infty\frac{d\xi}{2\pi}\ n_T(\xi)
\, i \, {\rm Tr}  \left[\ln(1-\mathbf{M}(i\xi))-\ln(1-\mathbf{M}(-i\xi))\right]
\end{equation}
is the temperature dependent part of the free energy containing the Boltzmann
factor $n_T(\xi)=1/(\exp(\xi/T)-1)$.

For the scalar field,  $\mathbf{M} (\xi)$ in (\ref{1.F1}) is a matrix in the orbital momentum
indices $l$ and $l'$ with matrix elements
\begin{equation}\label{1.M} M^{\rm  }_{l,l'}(\xi) =d^{\rm }_l(\xi R)
\sqrt{\frac{\pi}{4\xi L}}\sum_{l''=|l-l'|}^{l+l'}K_{\nu''}(2\xi
L)\,H_{ll'}^{l''}\,
 \,.
\end{equation}
Here, the function $d_l(x)$ results from the  T-matrix for the scattering on the sphere.
For Dirichlet and Neumann boundary conditions on the sphere we note
\begin{equation}\label{1.d}
d^{\rm  D}_l(x )=\frac{I_{\nu}(x)}{K_{\nu}(x )},
\quad
d^{\rm N}_l(x )=
\frac{\left(I_{\nu}(x)/\sqrt{x }\right)'}{\left(K_{\nu}(x )/\sqrt{x}\right)'}\,.
\end{equation}
In these formulas, $R$ is the radius of the sphere, $L$ is the separation between the
plane and the center of the sphere, $I_{\nu}(x)$ and $K_{\nu}(x)$ are the modified Bessel
functions.
We introduced the notations $\nu=l+1/2$, $\nu'=l'+1/2$ and $\nu''=l''+1/2$,
which will be used throughout the paper.

The factors $H_{ll'}^{l''}$
in (\ref{1.M}) result from the translation formulas. Their explicit form is
\begin{equation}
H_{ll'}^{l''}=  \sqrt{(2l+1)(2l'+1)}(2l''+1)
\left(\begin{array}{ccc}l&l'&l''\\0&0&0\end{array}\right)
\left(\begin{array}{ccc}l&l'&l''\\m&-m&0\end{array}\right),
\label{1.H} \end{equation}
where the parentheses denote the $3j$-symbols.

The above formulas are for Dirichlet boundary conditions on the plane.  For Neumann boundary conditions
on the plane we have to
reverse the sign in the logarithm in (\ref{1.F2}) or, equivalently, to change the sign of $\mathbf{M}$.
The trace in (\ref{1.F1}),
\begin{equation}\label{1.trace1} {\rm Tr}=\sum_{m=-l_m}^{l_m}\sum_{l=m}^{l_m} \ ,
\end{equation}
is over the orbital momenta truncated at some $l_m$. Of course, the final expression appears for $l_m\to\infty$.

For the electromagnetic field, the matrix $\mathbf{M}$ is in addition a matrix in the two polarizations.
These correspond to the TE and the TM modes in spherical geometry and we can represent the corresponding
matrix elements $\mathbb{M}_{ll'}$ as matrixes (2x2),
\begin{equation}
 \mathbb{M}_{l,l'}
=
\sqrt{\frac{\pi}{4\xi L}}\sum_{l''=|l-l'|}^{l+l'}K_{\nu''}(2\xi
L) H_{ll'}^{l''}\,
 \left(\begin{array}{cc}\Lambda_{l,l'}^{l''}&\tilde{\Lambda}_{l,l'}
\\ \tilde{\Lambda}_{l,l'}&\Lambda_{l,l'}^{l''}\end{array}\right)
\left(\begin{array}{cc}d^{\rm TE}_l(\xi R)&0
\\ 0&-d^{\rm TM}_l(\xi R)\end{array}\right)
 \,\label{1.M2}
\end{equation}
with the factors
\begin{equation}\label{2.LA}
\Lambda_{ll'}^{l''}=\frac{\frac12\left[l''(l''+1)-l(l+1)-l'(l'+1)\right]}
{\sqrt{l(l+1)l'(l'+1)}}\,,
\quad
\tilde{\Lambda}_{ll'}=\frac{2m\xi L}{\sqrt{l(l+1)l'(l'+1)}}\,,
\end{equation}
%
which follow from the translation formulas for the vector field. The factors resulting
from
the scattering T-matrices are
\begin{equation}\label{1.dTE}
 d^{\rm TE}_l(x)= \frac{I_{\nu}(x)}{K_{\nu}(x)}\,,
 \quad
 d^{\rm TM}_l(x)= \frac{\left(I_{\nu}(x)\sqrt{x}\right)'} {\left(K_{\nu}(x)\sqrt{x}\right)'}\,.
\end{equation}
When inserting these expressions into (\ref{1.F1}) or (\ref{1.F2}), the
trace must be taken also over the polarizations and the orbital momentum sum is restricted by $l\ge \max(1,|m|)$.

\section{The low temperature expansion}
Due to the Boltzmann factor in (\ref{1.F2}), the low temperature expansion emerges from the expansion, for $\xi\to0$, of
\begin{equation}\label{2.M1}
\mathbf{M}(\xi)=\mathbf{M}_0+\mathbf{M}_1\,(L\xi)^1+
\mathbf{M}_2\,(L\xi)^2+\mathbf{M}_3\,(L\xi)^3+\dots\,.
\end{equation}
The coefficients $\mathbf{M}_i=\mathbf{M}_i(\rho)$ are dimensionless functions of the ratio
\begin{equation}\label{2.ep}\rho=\frac{R}{L}.
\end{equation}
Inserting the expansion  (\ref{2.M1}) into the trace of the logarithm and keeping only the first two odd orders we get
\begin{equation}\label{2.M2}
{\rm Tr} \ln\left(1-\mathbf{M}(\xi)\right)=N_1(\rho)L\xi+N_3(\rho)(L\xi)^3+\dots
\end{equation}
with
\begin{eqnarray}\label{2.N}
N_1&=&-{\rm Tr}\left[ \left(1-M_0\right)^{-1}M_1\right] \,,\\\nonumber
N_3&=&-{\rm Tr} \left[\left(1-M_0\right)^{-1}M_3\right]\nonumber\\\nonumber
&&
-{\rm Tr} \left[\left(1-M_0\right)^{-1}M_1\left(1-M_0\right)^{-1}M_2\right]
-\frac{1}{3}\,{\rm Tr} \left[\left(\left(1-M_0\right)^{-1}M_1\right)^3\right]\,,
\end{eqnarray}
which are functions of $\rho$ like the $\mathbf{M}_i$'s.

It must be mentioned that inserting (\ref{2.N}) into (\ref{1.F2})
we interchange the orders of the limits $T\to0$ and $l_m\to\infty$.
Below we will see in which cases this is justified and in which it is not. With
the expansion (\ref{2.N}), the low-$T$ contributions to  the free energy (\ref{1.F2}) are
\begin{equation}\label{2.F}
\Delta_T {\cal
F}=-\frac{\pi}{6}N_1(\rho)\,LT^2+\frac{\pi^3}{15}N_3(\rho)\,L^3T^4
+\dots\,.
\end{equation}
The corresponding contributions to the force are
\begin{equation}\label{2.f}
\Delta_T f\equiv-\frac{{\rm d}}{{\rm d} L}\Delta_T{\cal F}
=\frac{\pi}{6}\frac{{\rm d}\, (LN_1(\rho))}{{\rm d}\,L}\,T^2-
 \frac{\pi^3}{15}\frac{{\rm d}\, (L^3N_3(\rho))}{{\rm d}\,L}\,T^4+\dots\,.
\end{equation}
As it will turn out there is only one contribution   $\sim T^2$ to the force (Section 4.2, below)
and in all other examples considered in this paper  the low-$T$ expansion starts from $T^4$. This is in agreement with the findings of \cite{webe1003.3420}. In order to compare the results we expand (\ref{2.f}) for small separation $a=L-R$,
\begin{equation}\label{2.c}
\Delta_T f=\left(c_2 R^3+c_3 a R^2\right)T^4+\dots\,,
\end{equation}
where the coefficients $c_2$ and $c_3$ were introduced in the same way as in \cite{webe1003.3420}

\section{Results for hard boundary conditions on the sphere}
In this section we consider hard boundary conditions on both, the plane and the sphere. We start with the scalar field. Here we have Dirichlet (D) or Neumann (N) boundary conditions and we denote the four combinations by (X,Y), where X stands for the sphere and Y stands for the plane. For example, (D,N) denotes Dirichlet boundary condition on the sphere and Neumann boundary conditions on the plane.
We remind that Neumann boundary conditions on the plane appear from reversing the sign of $\mathbf{M}$.

\subsection{The case DD}

\begin{figure}[b]
\epsfig{file=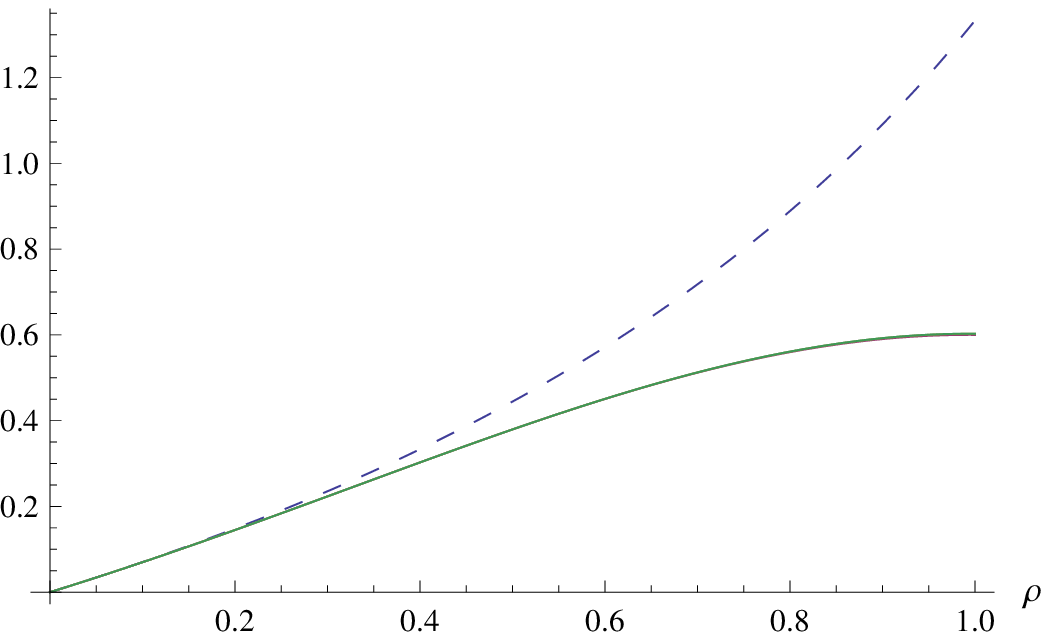,width=5.70cm}
\epsfig{file=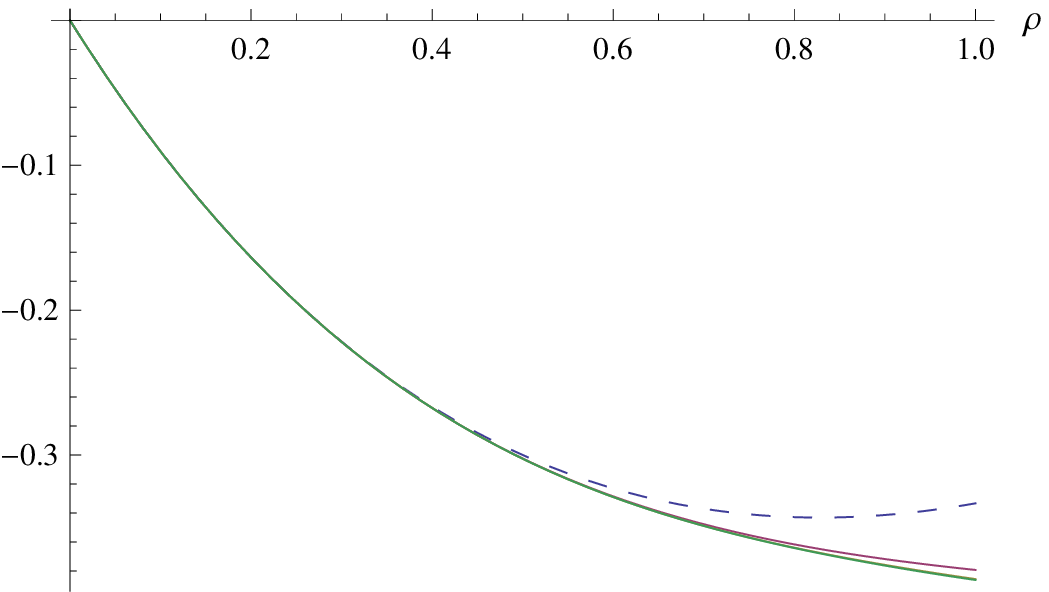,width=5.70cm} \caption{The functions
$N_3(\rho)$ for the case (D,D) (left panel) and $N_1(\rho)$ for the
case (D,N) (right panel)
 for several values of the truncation $l_m$. The limit of small separation corresponds
 to $\rho=1$. The dashed line  corresponds to $l_m=0$, i.e., to the pure s-wave contribution.
 Already for $l_m\ge1$ there is nearly no dependence on $l_m$.}
\label{figDD}
\end{figure}

In this case we have a non-zero
\begin{equation}\label{2.1.N1}
N_1=\rho\,,
\end{equation}
which is independent on the truncation $l_m$. This is the only case where one of the function $N_1$ or $N_3$
considered in this paper does not depend on the truncation.

The function $N_1$, (\ref{2.1.N1}), delivers the  $T^2$-contribution to the free energy which was found
in \cite{bord10-81-085023}. It does not contribute to the force since the dependence on $L$ drops out.
The next-order contribution is $N_3$. It is a rational function of $\rho$. The orders of the polynomials
in numerator and denominator grow with the order $l_m$ of the truncation. This is a general feature and holds
 for all functions $N_1$ and $N_3$ considered below except for those which vanish.

We display $N_3(\rho)$ as a function of $\rho$ in fig. \ref{figDD}
(left) for several $l_m$. It is seen that already for $l_m\ge1$ the
curves cease to change. In this way the free energy and the force
have a well defined limit for  $l_m\to\infty$. The coefficients
$c_2$ and $c_3$ defined in (\ref{2.c}) are shown in Table
\ref{tabDD} and it is seen that these fit well to those found in
\cite{webe1003.3420}.

At large separation, i.e., for a small sphere, only the lower  orbital momenta are on work.
The function $N_1$ is given by eq. (\ref{2.1.N1}), the function $N_3$ by
\begin{equation}\label{2.ls1}
N_3=\frac23 \, \rho+\frac13 \,\rho^2-\frac16 \,\rho^3+O(\rho^4)\,.
\end{equation}
In this way, the leading order temperature correction to the free energy and to the force do not depend on separation, while the subleading, $\sim T^4$, correction does.

\begin{table}[h]
\begin{center}
\begin{tabular}{c|ccccccccc}
\hline
$l_m$&0&1&2&3&4&5&6&7&8\\
\hline
$c_2$&-2.756&-3.748&-3.770&-3.772&-3.772&-3.772&-3.772&-3.772&-3.772\\
$c_3$&-5.512&-2.910&-2.500&-2.429&-2.426&-2.427&-2.426&-2.425&-2.425\\
\hline
\end{tabular}
\end{center}
\label{tabDD}
\caption{The values of the coefficients  $c_2$ and $c_3$ defined in eq. (\ref{2.c}) for the case (D,D)
 for several values of the truncation $l_m$. The corresponding values found in \cite{webe1003.3420},
 eq. (25), are $c_2=-3.96$ and $c_3=-2.7$.}
\end{table}


\subsection{The case DN}
In this case, as in the previous one, the dominating contribution is $N_1$. However, now it depends on the order $l_m$ of truncation. The first two orders are
\begin{equation}\label{2.DN1}
N_1(\rho)_{|_{l_m=0}}=\frac{\rho(-2+\rho)}{2+\rho}\,,
\quad
N_1(\rho)_{|_{l_m=1}}=\frac{\rho(-16+8\rho-4\rho^3+\rho^4)}{16+8\rho+4\rho^3+\rho^4}\,.
\end{equation}
%
In fig. \ref{figDD} (right panel), $N_1$ is shown as function of $\rho$ for several values of $l_m$. It is seen that there is a rapid convergence for large $l_m$. The sign is reversed as compared to the case (D,D) like for parallel plates.

Because of the more involved dependence on $\rho$ as compared to (\ref{2.1.N1}), $N_1$ contributes to the force. Hence, for these boundary conditions, according to (\ref{2.f}), we have a contribution  $\sim T^2$ to the force.

\subsection{The cases ND and NN}

\begin{figure}[h]
\epsfig{file=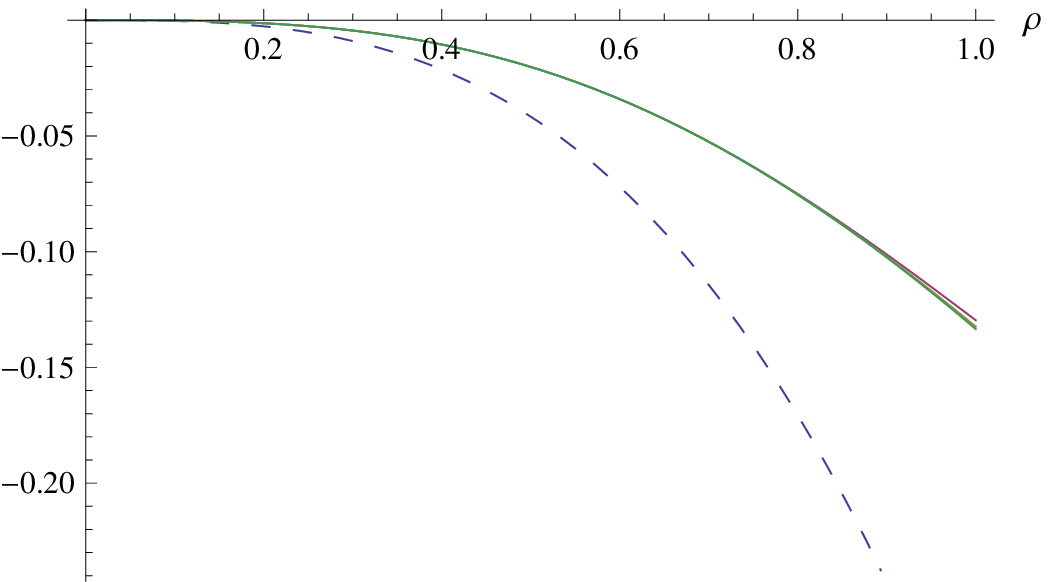,width=5.7cm}
\epsfig{file=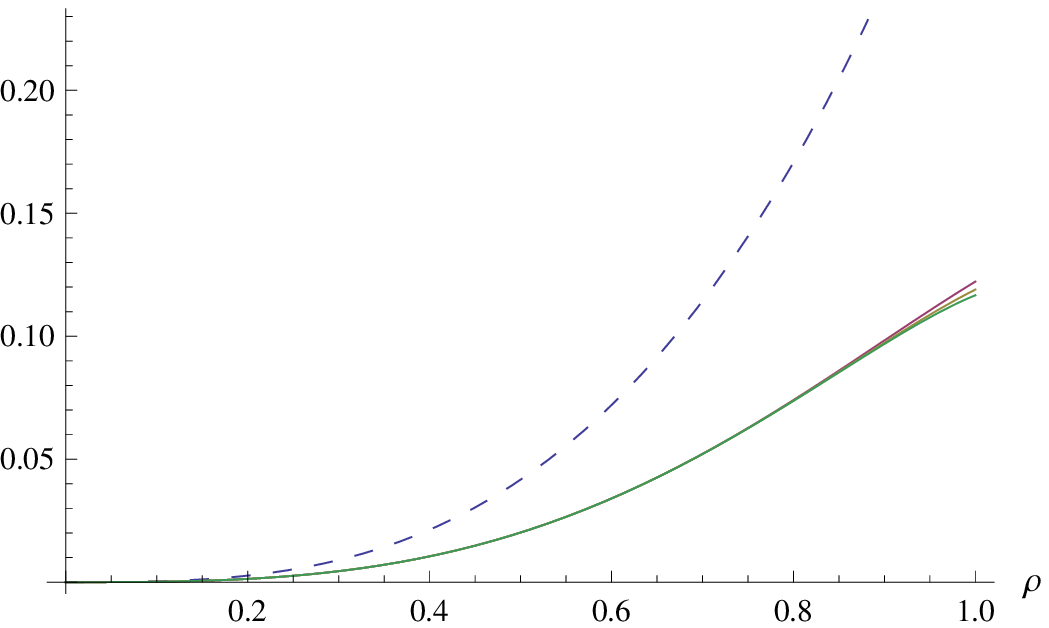,width=5.7cm}
\caption{The functions $N_3(\rho)$ for the cases (N,D) (left panel) and (N,N) (right panel) for several values of the truncation $l_m$. The limit of small separation corresponds to $\rho=1$. The dashed line  corresponds to $l_m=0$, i.e., to the pure s-wave contribution. Already for $l_m\ge1$ there is nearly no dependence on $l_m$.}
\label{figNX}
\end{figure}

In these two cases, which have Neumann boundary conditions on the sphere, the contribution of $N_1$ is zero  for all $l_m$  and the expansion starts with $N_3$.  These functions share the common features discussed above. We displayed both cases in fig. \ref{figNX}.

We mention that in \cite{bord10-81-085023} we considered only the
contribution from $l_m=0$ and missed the higher order terms.

The expansions for large separation is
\begin{eqnarray}\label{2.ls4}
N_3^{\rm (ND)}&=&-\frac16 \, \rho^3+\frac{1}{24} \,\rho^6-\frac{1}{192} \,\rho^9+O(\rho^{11})\,,
\nonumber\\
N_3^{\rm (NN)}&=&\frac16 \, \rho^3-\frac{1}{24} \,\rho^6-\frac{1}{384} \,\rho^9+O(\rho^{11})\,.
\end{eqnarray}
The first two terms are the same (up to the sign), higher orders are different.

\subsection{The electromagnetic field with conductor boundary conditions}
\begin{figure}[h]
\epsfig{file=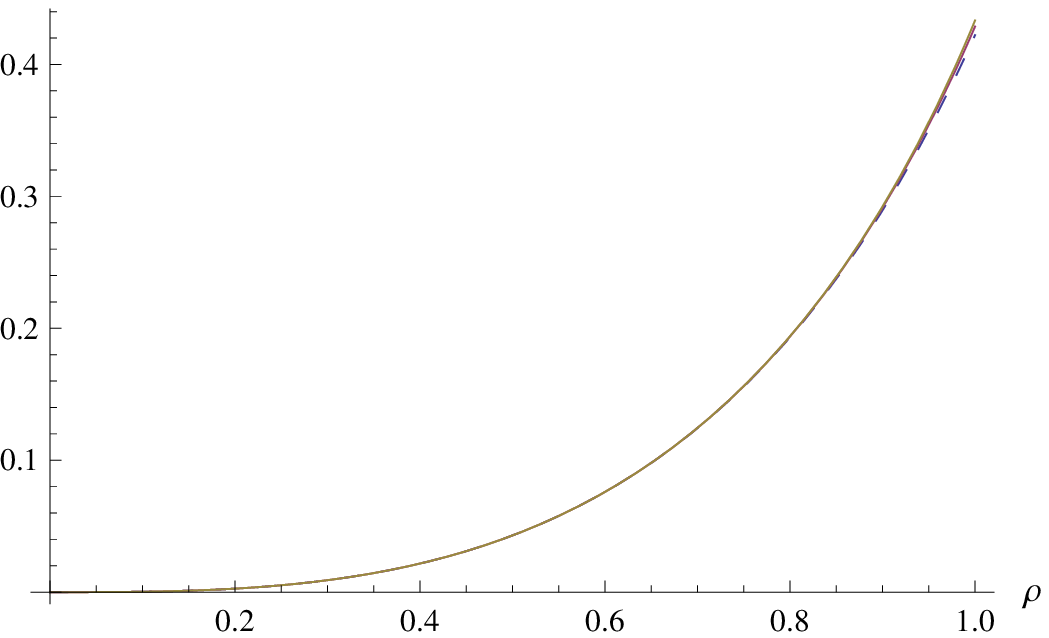,width=5.7cm}
\epsfig{file=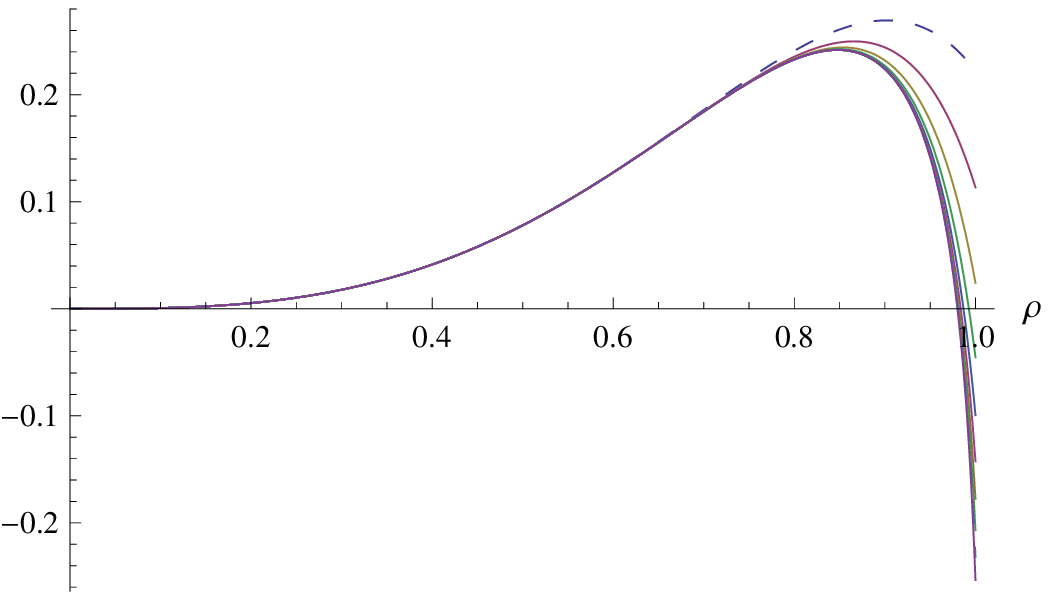,width=5.7cm}
\caption{The functions $N_3(\rho)$ for the electromagnetic field with conductor boundary
conditions for the TE polarization (left panel) and for the TM polarization (right panel)
for several values of the truncation $l_m$. The limit of small separation corresponds to $\rho=1$.
The dashed line  corresponds to $l_m=1$, i.e., to the pure p-wave contribution which is the lowest
one in the electromagnetic case. In the TE case, already for $l_m\ge1$, there is nearly no dependence
on $l_m$. For the TM case, at $\rho\lesssim 1$, there is no convergence for growing $l_m$.
We displayed until $l_m=10$. }
\label{figTX}
\end{figure}

For the electromagnetic field, the matrix elements $\mathbb{M}_{ll'}$ are given by eq. (\ref{1.M2}). We expand them in powers of $\xi$ as before and obtain an expansion in parallel to eq. (\ref{2.M1}),
\begin{equation}\label{2.M3}
\mathbb{M}(\xi)=\mathbb{M}_0+\mathbb{M}_1\,(L\xi)^1+
\mathbb{M}_2\,(L\xi)^2+\mathbb{M}_3\,(L\xi)^3+\dots\,.
\end{equation}
It turns out that the matrixes $\mathbb{M}_i$  are diagonal in the polarizations
for $i=0$ and $i=2$ and anti-diagonal for $i=1$.
Therefore in eq. (\ref{2.M2}), the first contribution, $N_1$, vanishes and from $N_3$ only
the term in eq. (\ref{2.N}) does not vanish. From this structure it follows also that the
contributions from the two polarizations do not mix in $N_3$ and we can consider these separately.
Now the further calculations go in the same way as for the scalar field and we have calculated
the functions $N_3(\rho)$ for both polarizations. We displayed them in fig. \ref{figTX}
While the TE mode gives a result similar to the scalar case in the sense that the
limit $l_m\to\infty$ is approached very fast, the picture for the TM mode is different. Here we observe, for $\rho$ close to unity, $\rho\lesssim1$,    contributions growing with $l_m$. This must be interpreted as a noncommutativity of the limits $T\to0$ and $l_m\to\infty$. As a consequence, at small separation, we have to expect contributions decreasing for $T\to0$ slower than $T^4$.

For large separation we find the following expansions,
\begin{eqnarray}\label{2.ls5}
N_3^{\rm TE}&=&\frac13 \, \rho^3+\frac{1}{12} \,\rho^6+\frac{1}{192} \,\rho^9+O(\rho^{11})\,,
\nonumber\\
N_3^{\rm TM}&=&\frac23 \, \rho^3-\frac{1}{3} \,\rho^6-\frac{1}{12} \,\rho^9+O(\rho^{11})\,.
\end{eqnarray}
According to (\ref{2.F}), the leading order contribution to the free energy is
\begin{equation}
\Delta_T {\cal F}=\frac{\pi^3}{15}R^3 T^4
-\frac{\pi^3}{60}\frac{R^6}{L^3}T^4+\dots
\end{equation}
and we observe a $T^4$-contribution to the force (from the second term). The first term coincides with the corresponding term  in eq. (6) in \cite{cana10-1005.4294} while the second is beyond  of what is displayed there.


\section{Results for a dielectric ball in front of a conducting plane}
For a dielectric ball the formulas of section 2 remain valid except for the functions $d_l^{\rm TX}(x)$, eq. (\ref{1.dTE}).
These must be substituted by
\begin{equation} \label{3.dTE}
d^{\rm TE}_l(z)=\frac{2}{\pi}\,\frac
{ \sqrt{\varepsilon}\,s_l(x)s_l'(nx)-\sqrt{\mu}\,s_l'(x)s_l(nx)  }
{\sqrt{\varepsilon}\, e_l(x)s_l'(nx)-\sqrt{\mu}\,e_l'(x)s_l(nx) }
\end{equation}
with the refraction index $n=\sqrt{\varepsilon\mu}$ and $s_l(x)=\sqrt{\pi x/2}\,I_\nu(x)$ and $e_l(x)=\sqrt{2 x/\pi}$ $K_\nu(x)$ are the modified spherical Bessel functions. The function $d^{\rm TM}_l(z)$ can be obtained from (\ref{3.dTE}) by interchanging $\varepsilon$ and $\mu$. 

Inserting these formulas into (\ref{1.M2}) and calculating the entries in eq. (\ref{2.M3}) we see that the structure of the  matrixes $\mathbb{M}_i$ remains the same. In this way we have a separation into TE and TM modes as before. As a consequence, we have   only   $T^4$ contributions.

Now we calculate the function $N_3(\rho)$ for the case of a fixed $\varepsilon$ and
for an $\varepsilon$ taken from the plasma model.

\subsection{Fixed permittivity $\varepsilon$}
\begin{figure}[h]
\epsfig{file=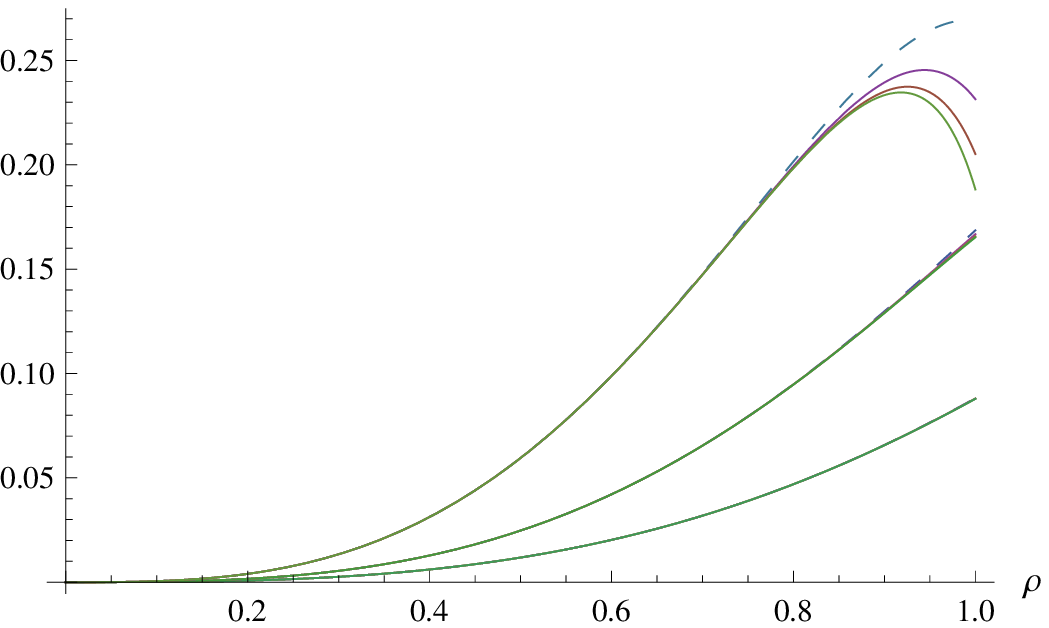,width=5.7cm}
\epsfig{file=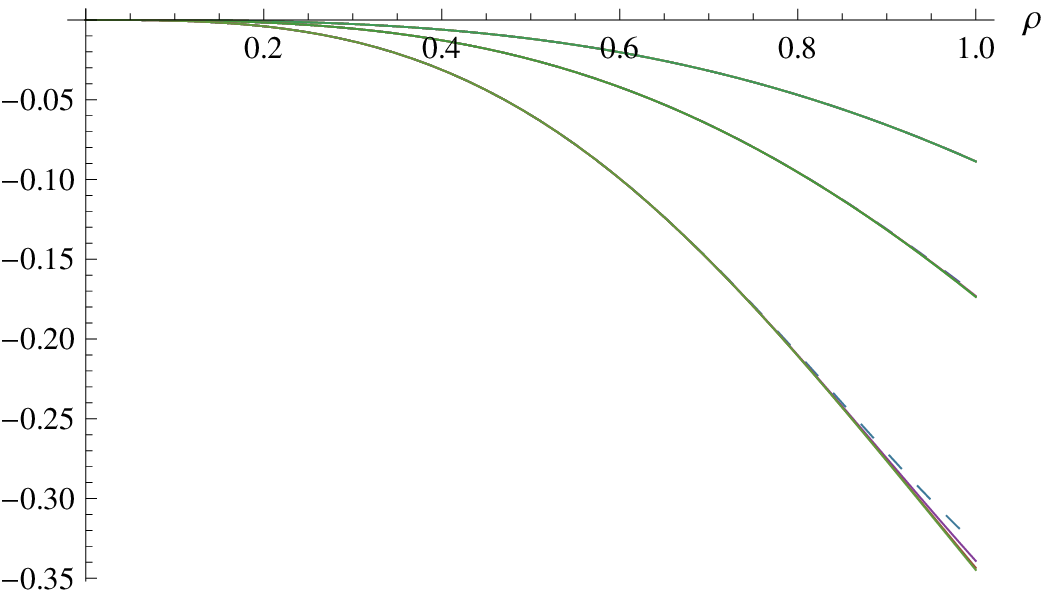,width=5.7cm}
\caption{The functions $N_3^{\rm TM}(\rho)$ for the dielectric ball (left panel) with $\mu=1$,
$\varepsilon=1.5$ (lower curve), $\varepsilon=2.3$ and $\varepsilon=10$,
and $N_3^{\rm TE}(\rho)$ (right panel) and with $\varepsilon=1$, $\mu=1.8$  (upper curve), $\mu=2.3$ and $\mu=10$. The dashed line is for $l_m=1$.  }
\label{figepTX}
\end{figure}

For fixed $\varepsilon$ we consider two cases. First we put $\mu=1$. In this case it turns out
that  $N_3(\rho)=0$ for the TE mode. This means that the corresponding low-$T$ expansion starts from
a higher power in $T$ which we do not consider here. For the TM mode the function $N_3(\rho)$
is shown in fig. \ref{figepTX} (left panel). It depends on the truncation. For $\epsilon$ close to unity it stabilizes rapidly, for higher $\epsilon$ slower.

At large distances, $\rho\to 0$, we found
\begin{equation}
N_3^{\rm TM}=\frac{2 (\epsilon -1)}{3 (\epsilon +2)}\,  \rho ^3-\frac{(\epsilon -1)^2 }{3 (\epsilon +2)^2}\, \rho ^6-\frac{(\epsilon -1)^3 \rho ^9}{12 (\epsilon +2)^3}\,  \rho ^9+O\left(\rho ^{11}\right)
%
\label{N3diel_large}
\end{equation}
In dilute approximation, $\epsilon=1+\delta$, $\delta\ll 1$, only the lowest orbital momenta contribute until the order quadratic in $\delta$,
\begin{equation}
N_3^{\rm TM}=\frac{2}{9}\rho^3 \delta-\frac{1}{27}\rho^3
(2+\rho^3)\delta^2+\mathcal{O}(\delta^3)\,  ,\label{N3dilute}
\end{equation}
higher orders are more complicated to obtain.

As the second case we consider $\varepsilon=1$. Here the
contribution of the TM mode to $N_3(\rho)$ is zero and we are left
with the TE contribution. This function is very similar to that in
the first case, however, different in details. It is shown in
fig. \ref{figepTX} (right panel). It stabilizes much faster when lifting the truncation
as in the previous case.
 For large separations it reads
\begin{equation}
N_3^{\rm TE}=-\frac{2( \mu-1 ) }{3 (\mu +2)}\,\rho ^3+\frac{(\mu -1)^2 }{3 (\mu +2)^2}\,\rho ^6-\frac{(\mu -1)^3 }{24 (\mu +2)^3}\,\rho ^9+O\left(\rho ^{10}\right)\,.
\end{equation}
The difference (up to the sign) starts with order $\rho^9$.
In dilute approximation we found in the first two orders the same expression as in eq. (\ref{N3dilute}) with reversed sign. Differences show up starting from the third order.

\subsection{Plasma model permittivity}
\begin{figure}[h]
\epsfig{file=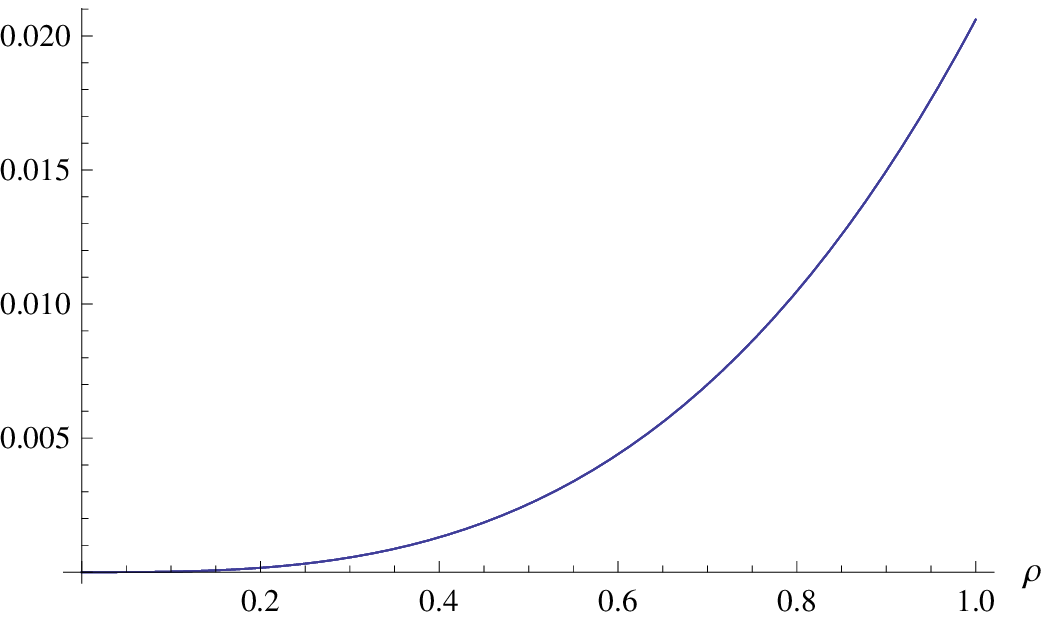,width=5.7cm}
\epsfig{file=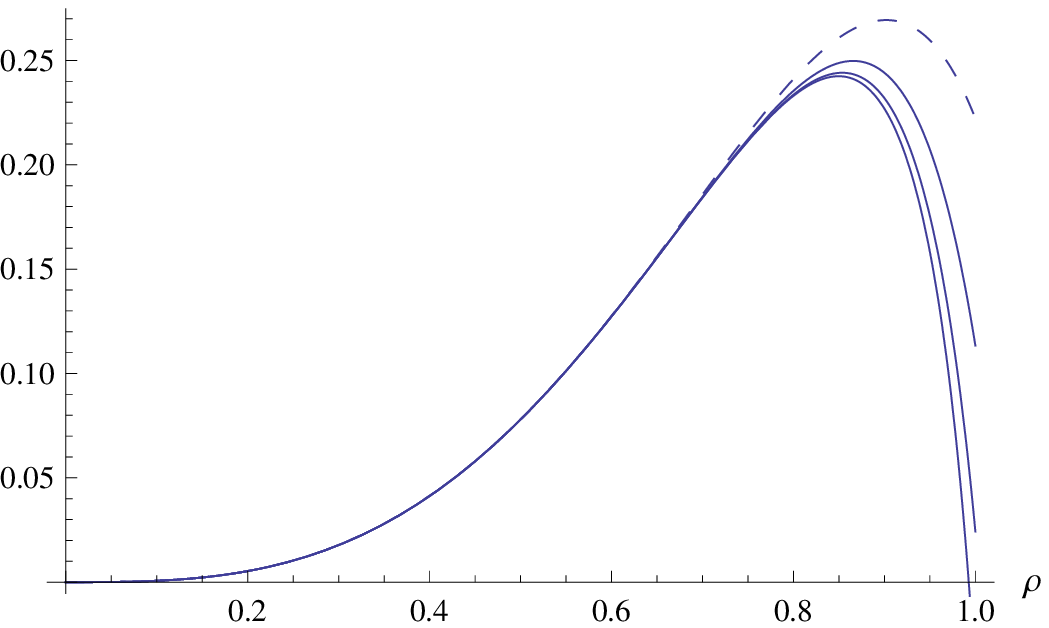,width=5.7cm}
\caption{The functions $N_3(\rho)$ for the dielectric ball with dispersion (\ref{epplasma})
of the plasma model for $\omega_p=1$. For the TE mode (left panel) the convergence for $l_m\to\infty$ is rapid for
all separations. For the TM mode (right panel), there is no convergence for $l_m\to\infty$ at small separation,
i.e., for $\rho\lesssim 1$. The dashed curve corresponds to $l_m=1$.}
\label{figDisp}
\end{figure}

The permittivity derived within the plasma model for metals is
\begin{equation}\label{epplasma}\varepsilon=1+\frac{\omega_p^2}{\xi^2}\,,
\end{equation}
where $\omega_p$ is the plasma frequency and $\varepsilon$ is taken for imaginary frequency $\xi$. Inserting (\ref{epplasma}) into (\ref{3.dTE}), the calculation runs as in the previous cases.
The results are shown in fig. \ref{figDisp}

For the TE mode, the curves cease to change already for $l_m=2$ within the precision of the plot. The analytic expressions are  rational function of $\rho$  and of hyperbolic functions of $\omega_p$. For small $\omega_p$ we observe
\begin{equation}\label{3.dispTE}N_3^{\rm TE}(\rho)=\frac{\rho^3\omega_p^2}{45}+O(\omega_p^3)\,.
\end{equation}
A different picture we observe for the TM mode. Here the truncation can be removed for $\rho<1$ only. At close separation, i.e., for $\rho\lesssim 1$, the contributions do not tend to a finite limit for $l_m\to\infty$. Again, we have to interpret this a non-commutativity of the limits $t\to0$ and $l_m\to\infty$. Hence we expect for small separation lower powers in $T$.
Also the behavior for $\omega_p\to0$ is nonanalytic in the sense, that $N_3(\rho)^{\rm TM}$ does not vanish in this limit,
\begin{equation}\label{3.dispTM}N_3^{\rm TM} (\rho)=\frac{\rho^3(-4+3\rho^2)}{3(-4+\rho^3)}+O(\omega_p)\,.
\end{equation}

At large distances, the leading correction to the free energy for the
TM mode does not depend on the plasma frequency,
$$N_3^{\rm TM}=\frac{2 \rho^3}{3}+{\cal O}(\rho^5).$$
While for the TE mode the correction is sensitive to small plasma
frequencies,
$$N_3^{TE}=\left(\frac{1}{3}+\frac{1}{\omega_p^2}-\frac{\coth(\omega_p)}{\omega_p}\right)
\rho^3 +{\cal O}(\rho^5)\,,$$ but saturates at $1/3$ when
$\omega_p\to\infty$.
\section{Conclusions}
In the forgoing sections we calculated the low temperature expansion of the free energy for a sphere
or a dielectric ball in front of a plane. For the temperature dependent part $\Delta_T {\cal F}$ of
the free energy we used the representation (\ref{1.F2}) involving the Boltzmann factor. Further we used
a truncation of the orbital momentum sum, $l\le l_m$, and interchanged the limits $T\to0$ and $l_m\to\infty$.
After that, the low-$T$ expansion is obtained simply by expanding the matrices $\mathbf{M}$ into powers of $\xi$
and taking the lowest odd one. In this way, the low-$T$ expansion takes the generic form
\begin{equation}\label{6.F}
\Delta_T {\cal F} ={\cal F}_2 T^2+{\cal F}_4 T^4+\dots\,.
\end{equation}
The coefficient ${\cal F}_2$ is present for the cases (D,D), section 4.1 (but independent on the separation)
and (D,N), section 4.2. It is zero in all other cases where ${\cal F}_4$ is the leading order contribution.

For all examples considered in this paper, at finite separation, ${\cal F}_2$  and ${\cal F}_4$ have
a finite limit for $l_m\to\infty$. Hence the generic low-$T$ behavior is given by eq. (\ref{6.F}). This holds, for instance, at large separation.
A different picture appears for small separation, $\rho\to1$. In some
cases, the closer the separation, the worse the  convergence for $l_m\to\infty$.
In these cases we do not have a result for $T\to0$. However, we can expect lower powers of $T$ to appear. This is a topic of future investigations.


\begin{acknowledgement}
This work was supported by the Heisenberg-Landau program. The
authors benefited from exchange of ideas by the ESF Research Network
CASIMIR. I.P. acknowledges partial financial support from FRBR
grants  09-02-12417-ofi-m and 10-02-01304-a.
\end{acknowledgement}

\end{document}